\documentclass[aps,prl,twocolumn,notitlepage,letterpaper]{revtex4-2}
\usepackage[utf8]{inputenc}
\usepackage{graphics,amssymb,amsmath,epsfig,color}
\usepackage{graphicx}
\usepackage{float}
\usepackage{braket}
\usepackage{mathptmx}
\usepackage{times}
\usepackage{mathtools}
\usepackage{bm}
\usepackage{ifpdf}

\newcommand{\bA}{{\mathbf A}}

\newcommand{\bE}{{\mathbf E}}

\newcommand{\bz}{{\mathbf z}}

\usepackage{amsfonts}

\usepackage{amsmath}
\usepackage{amssymb}
\usepackage{bm}
\usepackage{mathrsfs}
\usepackage{tensor}
\usepackage{dsfont}
\usepackage{graphicx}
\usepackage[caption=false]{subfig}

\usepackage[usenames,dvipsnames]{xcolor}
\usepackage[hyperindex,pdftex, breaklinks,colorlinks = true,linkcolor = blue,urlcolor=blue,citecolor=blue]{hyperref}

\usepackage{physics}

\begin{document}
\title{Capacitive Scheme to Detect the Topological Magnetoelectric Effect}
\author{Chao Lei$^{1}$}
\email{leichao.ph@gmail.com}
\author{Perry T.~Mahon$^{1}$}
\author{C.~M.~Canali$^{2}$}
\author{A.~H.~MacDonald$^{1}$}
\affiliation{$^{1}$Department of Physics, University of Texas at Austin, Austin, Texas 78712, USA}
\affiliation{$^{2}$ Department of Physics and Electric Engineering, Linnaeus University, 392 31 Kalmar, Sweden}

\date{\today}

\begin{abstract}
The topological magnetoelectric effect (TME) is a defining property of three-dimensional $\mathbb{Z}_{2}$ topological 
insulators that was predicted on theoretical grounds more than a decade ago, but has still not been directly measured.
In this Letter we propose a strategy for direct measurement of the TME and discuss the precision 
of the effect in real devices with charge and spin disorder.
\end{abstract}
\maketitle

\textit{Introduction---}
The electronic ground state of bulk three-dimensional (3D) time-reversal invariant band insulators can be 
classified by a $\mathbb{Z}_2$-valued index \cite{TME_TI_2008,Qi_review_2011,Hasan_review_2010,Mong2010,monaco2015symmetry}. In thin films of $\mathbb{Z}_2$-odd materials, the nontrivial bulk topology implies the existence of
an odd number of surface state Dirac cones that are commonly revealed
by angle-resolved photoemission spectroscopy \cite{Hsieh2008,Xia2009,Chen2009}.  
One key property \cite{TME_TI_2008,Vanderbilt2009} of bulk 3D $\mathbb{Z}_2$ topological insulators (TIs) is
the quantization of their formal magnetoelectric response coefficient.  The formal bulk response 
manifests as an observable topological magnetoelectric effect (TME) 
only when the surface states are gapped by locally breaking time-reversal symmetry so that 
the Dirac cones support a nonzero surface Hall conductivity \cite{TME_TI_2008,Vanderbilt2009,Essin2010,Maciejko2010,souza2011chern,Nomura2011,Rosenow2013,
Burnell2013,Sekine_2021,ahn2022theory,Nagaosa2015,Witten2016,mahon2023reconciling}.  
Although first proposed more than 15 years ago, the TME has not yet been directly measured.
In this Letter we discuss why the observation of the TME has been difficult, propose a strategy for its 
measurement, and speculate on the precision of the effect's quantization in real samples with disorder.

Magnetoelectric response \cite{fiebig2005revival} refers to the linear response of (electronic) polarization $\bf{P}$ in a dielectric to a uniform dc magnetic field ${\bf B}$, or 
equivalently to the linear response of orbital magnetization ${\bf M}$ to a uniform dc electric field ${\bf E}$, and is described by the susceptibility tensor
\begin{align}
\alpha^{il}=\left.\frac{\partial P^i}{\partial B^l}\right|_{\substack{\textbf{E}=\textbf{0} \\ \textbf{B}=\textbf{0}}} = \left.\frac{\partial M^l}{\partial E^i}\right|_{\substack{\textbf{E}=\textbf{0} \\ \textbf{B}=\textbf{0}}}.    
\end{align}
The TME of a ($\mathbb{Z}_{2}$-odd) TI
implies a response tensor that is diagonal and quantized: $\alpha^{il} = \delta_{i,l} (2n+1) e^2/2hc$  for surfaces with $2n+1$ Dirac cones ($n\in\mathbb{Z}$).
For concreteness, we focus below on TI thin films with one Dirac cone at each surface ($n=0$), as realized, for example, in Bi$_{2}$Te$_{3}$, 
but our conclusions are valid more generally.
In realistic materials, the TME is accurately quantized only when the sample's volume is large \cite{Nagaosa2015,Pournaghavi2021,mahon2023reconciling}. 
Because the TME has never been measured directly,
little is known with confidence about how it is affected by the inevitable disorder present in realistic devices.
 
\begin{figure}[t!]
    \centering
    \includegraphics[width=0.45\textwidth]{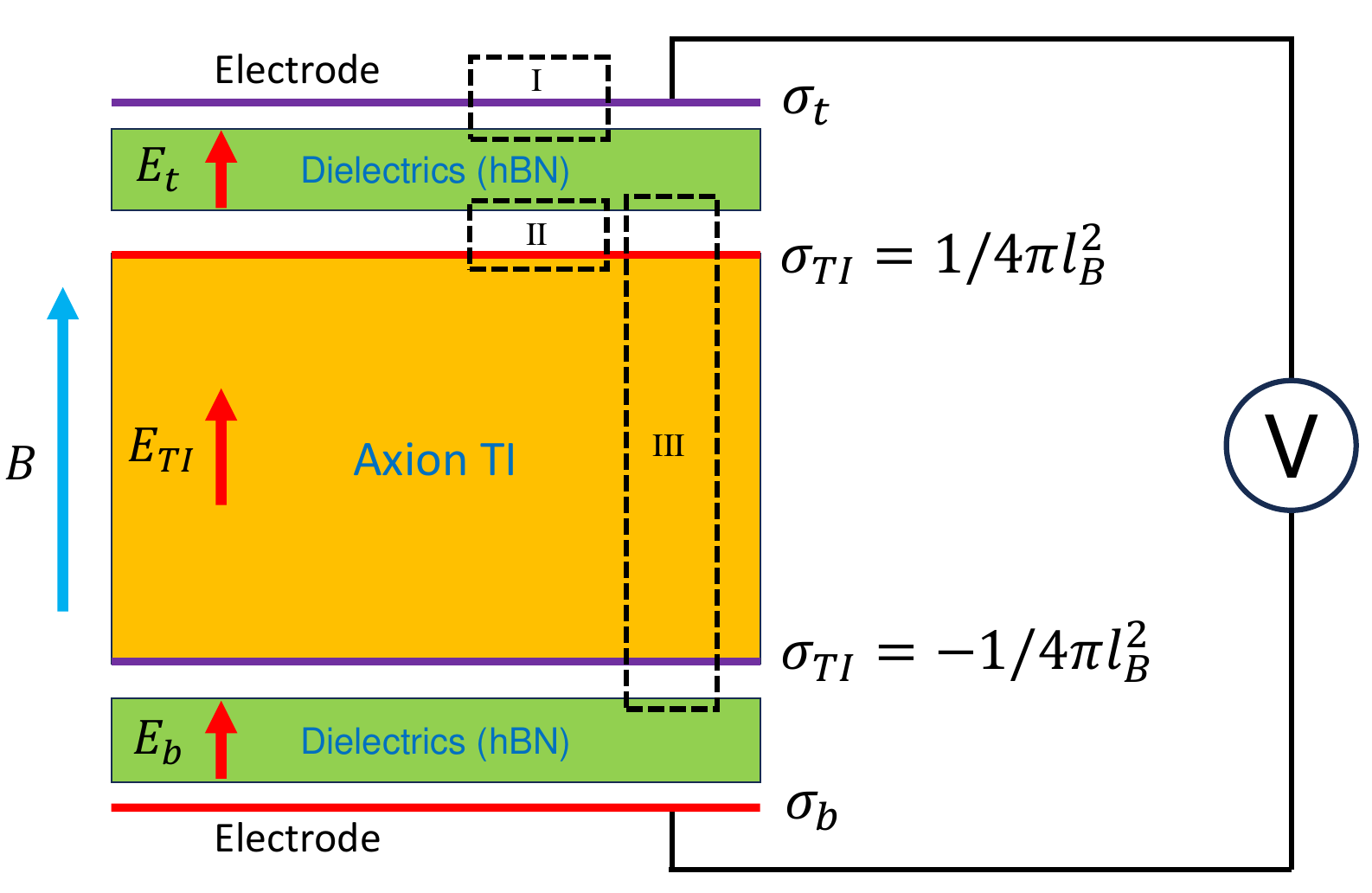}
    \caption{Measurement scheme for the magnetoelectric coefficient $\alpha$ of $\mathbb{Z}_{2}$-odd TI thin films.
    A floating nonmagnetic TI film with surface dopant magnetizations in an axion configuration (see main text) 
    is encapsulated by hexagonal boron nitride ($h$-BN) and placed between two parallel electrodes.
    The van der Waals stack therefore acts as a compound dielectric in a planar capacitor.
    When a magnetic field is applied in the surface normal direction, 
    the topological magnetoelectric effect
    is manifested by the accumulation of charges on the TI's two surfaces with  
    areal densities $\pm 1/4\pi l_B^2$
    (assuming there is a single Dirac cone at each surface), where $l_B$ is the magnetic length.
    The TI surface charge densities add an offset of $\pm C_{\text{TI}}^{-1} A \alpha_{me} B$ to the capacitance voltage at fixed electrode charge.
    The sign of the surface densities is related to the magnetic configuration at each surface, and for concreteness in this figure we have assumed magnetic surface dopants 
    in the out-of-surface configuration.}
    \label{fig:scheme}
\end{figure}

In this Letter we propose a simple strategy for measuring the 
TME of a TI thin film with nonmagnetic bulk and magnetic dopants at its top and bottom surfaces 
\footnote{In principle, it is fundamentally equivalent to consider a nonmagnetic TI thin film with 2D ferromagnetic insulator layers at its top and bottom surfaces. However, in present samples the exchange splitting of the nonmagnetic film's gapless surface states due to those ferromagnetic layers is weak \cite{Bhattacharyya2021,Mogi2019,Tand2017}. Any disorder in these samples will therefore significantly affect the quantization of the TME and indeed will greatly suppress magnetoelectric response overall, nullifying our proposed detection scheme. However, materials advances could make platforms of this type viable in the future.}.
In the thin film geometry, the TME occurs
only when the film's two surfaces have opposite magnetization orientations.  This enabling configuration of the 
TI thin film magnetization is often referred to as the axion configuration \cite{TME_TI_2008,Armitage2019,Sekine_2021} because of an analogy between
the electrodynamics implied by the TME and axion field theories \cite{Wilczek1987}.
When the two surface magnetizations are parallel 
\footnote{In this work we consider top and bottom surface magnetic dopant configurations with magnetization that is in the surface-normal direction.}, 
the system hosts a quantum anomalous Hall effect \cite{Chang_2013} 
and has chiral edge states which cross the Fermi level and violate the fixed charge condition needed
to define polarization in a dielectric.  
The TME measurement scheme we propose employs a floating sandwich structure, with two magnetic-dopant-free nonmagnetic $\mathbb{Z}_{2}$-even dielectric layers (e.g., hexagonal boron nitride ($h$-BN)) separating the TI sample from gates as shown in Fig.~\ref{fig:scheme}. 
An external uniform dc magnetic field with magnitude $B$ that is parallel to the surface normal of the thin film
induces a polarization in the same direction and manifests as an accumulated electron
density $\pm 1/4\pi l_B^2$, where $l_B^2=\hbar c/eB$ is the magnetic length, at the top and bottom surfaces of an axion configured TI.
This can be detected 
by measuring the magnetic field dependence of the capacitor voltage at fixed charge, as we explain below. 

In principle, any $\mathbb{Z}_{2}$-odd TI can support the TME, including TIs
with magnetic order in their bulk \cite{tokura2019magnetic}.  However, the measurement we 
propose relies on weak magnetic exchange coupling between top and bottom surfaces.
In antiferromagnetic TIs (e.g., MnBi$_{2}$Te$_{4}$), the magnetic order response to 
external magnetic fields is more complex (than in nonmagnetic axion TIs) because of 
strong interlayer exchange interactions, and 
is characterized by intermediate spin-flop configurations and complex switching patterns in interior layers
\cite{Lei2021}. If the goal is to measure the TME, these intermediate magnetic configurations 
are an unwelcome complication.

\textit{Capacitive TME measurement scheme---} 
As illustrated in Fig.~\ref{fig:scheme}, when a uniform dc surface normal magnetic field ${\bf B}$ is applied to a TI thin film with nonmagnetic bulk and an axion 
magnetic surface dopant configuration, electric polarization is induced and manifested by surface charge densities of opposite sign on the TI's top and 
bottom surfaces.
Electrical measurement of these surface charge densities is subtle because the currents that 
cause those densities to accumulate cannot be measured externally \cite{Nagaosa2015}.
Instead, the surface charges result from internal currents that flow as the parameters of an isolated equilibrium system are varied (see below) \cite{Essin2010}. 
Our goal in the following is to 
explain an electrical measurement scheme that can verify quantization if it is realized.
The basic idea is to avoid direct electrical contact to the TI, which sits in the interior 
of a capacitor's dielectric stack, in order to ensure its total charge neutrality (see Fig.~\ref{fig:scheme}). 
We apply Gauss's law 
\begin{equation}
    \oint_S \epsilon \bE \cdot d \bA = 4 \pi Q_{S},
\end{equation}
where $Q_{S} = e A_{S} \sigma_{S}$ is the total {\it free} charge 
\footnote{Free excludes charges associated with dielectric 
discontinuities but includes the magnetic field induced charge associated with the TME.} 
enclosed by a general surface $S$ with area $A_{S}$ and $e$ is the elementary charge,
to the Gaussian surfaces in Fig.~\ref{fig:scheme}.  
Taking $\hat{\bz}$ aligned normal to the top surface, by usual electrostatics arguments $\bE=E\hat{\bz}$ and application of Gauss's law to box {\it I} yields 
\begin{equation}
    E_t =  - 4 \pi e \sigma_t /\epsilon_{t}
\end{equation}
for the electric field in the top dielectric layer, where $e\sigma_t$ is the charge density on the top electrode and $\epsilon_{t}$ is the dielectric constant of the top dielectric material (e.g., $h$-BN).  When we apply Gauss's law to box {\it II}, 
which includes the top surface of the axion configured TI, we obtain
\begin{equation}
    \epsilon_{t} E_t - \epsilon_{\text{TI}}E_{\text{TI}} = 2\pi e  (1+\delta) B/\Phi_0.
\end{equation}
We have used $P=(eA\sigma_{\text{TI}})d_{\text{TI}}/V_{\text{TI}}$, where $V_{\text{TI}}=d_{\text{TI}}A$ is the volume of the TI film and $A$ is the cross-sectional area of the device,
and also $P=\alpha_{me}B$, where $\alpha_{me} = (e/2)(1+\delta)/\Phi_0$ 
and $\Phi_0=hc/e$.
$\delta$ is included to account for any deviations of the  
magnetoelectric response from the expected quantized value, which could be produced by disorder as we discuss below.
Finally, applying Gauss's law to box {\it III} implies that the electric fields in the top and bottom dielectrics are 
identical, as required by the charge neutrality of the floating TI film.  
The total voltage drop across the dielectric stack is therefore 
$V = - (E_{t} d_t + E_{b} d_b + E_{\text{TI}} d_{\text{TI}})$, which we write as 
\begin{equation}\label{eq:voltageA}
   V = s \, C_{\text{TI}}^{-1} A \alpha_{me} B + C_g^{-1} Q.
\end{equation}
Here
$\pm Q$ is the electric charge of the top and bottom electrode, respectively, $|s|=1$ in axion insulator magnetic configurations
and changes sign between upsweep and downsweep (see below), 
whereas $s=0$ in quantum anomalous Hall (QAH) magnetic configurations. 
In Eq.~(\ref{eq:voltageA}), $C_g^{-1} \equiv 4\pi \sum_j d_j/(\epsilon_j A)$ is the total geometric capacitance, $d_{j}$ is the thin film thickness of layer $j \in \{t,b,\text{TI}\}$ in the stack, and 
$C_{\text{TI}}^{-1} \equiv 4\pi d_{\text{TI}}/(\epsilon_{\text{TI}} A)$ is the geometric capacitance of the TI. 
The topological magnetoelectric coefficient can therefore be extracted by measuring the voltage across the device 
as an external magnetic field that is parallel to the surface normal is varied. 

\begin{figure}[t!]
    \centering
    \includegraphics[width=0.45\textwidth]{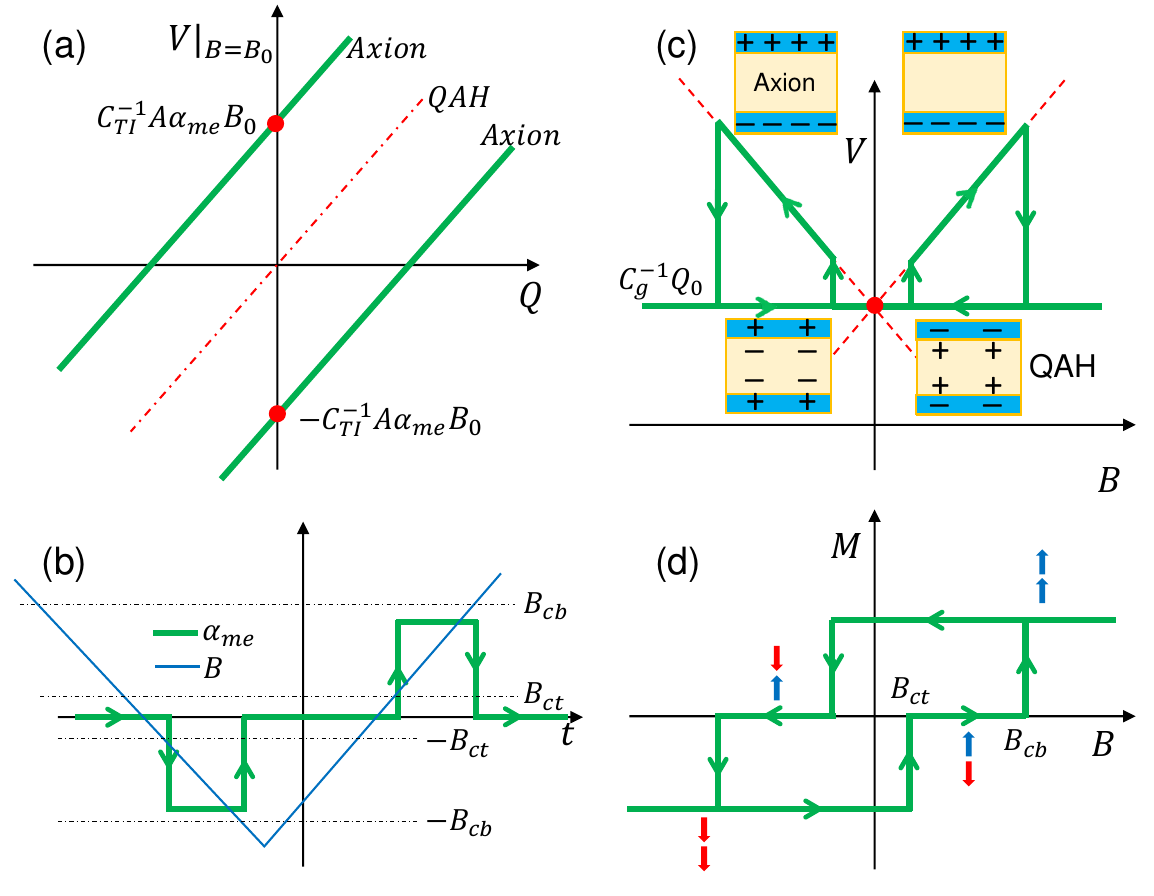}
    \caption{
    (a) Dependence of $V|_{B=B_{0}}$ on electrode charge $Q$ in QAH and axion insulator magnetic surface dopant configurations.
    The top (bottom) curve applies to axion configurations with surface magnetic moments parallel
    (antiparallel) to the surface normal. The middle curve applies to QAH configurations.
    (b) Time $t$ variation of magnetic field (blue curve) in the proposed hysteresis loop and the magnetoelectric coefficient (green curve) of the electronic state (QAH or axion) 
    expected at time $t$ when the two surface magnets switch at different 
    coercive fields $B_{ct} < B_{cb}$ and the field is swept slowly.      
    (c) Voltage drop across the dielectric stack during the proposed hysteresis loop at a fixed electrode charge $Q_0$ (described in main text).
    (d) Surface magnetic configuration during the proposed hysteresis loop. Because the top surface coercive field is assumed to be smaller, the axion configurations that appear in the upsweep and downsweep 
    branches are respectively out of surface ($\alpha>0$) and into surface ($\alpha<0$).}
    \label{fig:measurement}
\end{figure}

\textit{Capacitive hysteresis loop---} In Fig.~\ref{fig:measurement} we illustrate how the capacitor's voltage varies 
at fixed charge across a perpendicular field magnetic hysteresis loop.  The geometric capacitance and electrode area are 
device-dependent quantities. The dependencies of capacitance 
voltage on electrode charge when the external magnetic field is fixed at $B_0$ in axion insulator and 
QAH states are illustrated in Fig.~\ref{fig:measurement}(a).
Here all curves have slope $C_g^{-1}$ and the TME is manifested by offset voltages $\pm C_{\text{TI}}^{-1} A B_0 \alpha_{me}$
in the two axion magnetic configurations.  The TME is reflected most directly when $B$ is varied across a 
hysteresis loop.  We assume that the magnets formed by the two surface dopant systems are weakly
exchange coupled and that they have different coercive fields.  We start the hysteresis loop
from positive fields $B$ that exceed both coercive fields $B_{c,t/b}$ such that the magnetizations on top and 
bottom surfaces are aligned and parallel to the top surface normal. 
Thus, the system begins in a QAH state. 
We then follow the 
field loop summarized in Fig.~\ref{fig:measurement}(b). 
According to the Streda formula \cite{Streda1982,Streda1983,Widom1982,MacDonald1984}, the charge density at which the gap appears
is nonzero at finite magnetic field in a QAH state.  
The thin film, which is neutral, must therefore 
have a Fermi level outside 
the gap and behave as a metal in the sense of fully screening the external field deep in its bulk, reducing its contribution to 
the voltage compared to the dielectric contribution contained in $C_{g}^{-1}$.
This reduction, which we do not expect to be a large contribution to the total voltage, will
decrease in magnitude as the magnetic field strength is reduced toward zero.
For a sufficiently negative $B$ one of the two surface magnetizations 
($t$ for instance, if $B_{ct}<B_{cb}$) will 
flip, converting the magnetic configuration from QAH to axion.  
It is at this point that the topological insulator
surfaces first develop the opposite charge densities related to the 
TME, and these make the contribution to the voltage proportional to $\alpha_{me} B$ (see Fig.~\ref{fig:measurement}(c)).  
At stronger negative values of $B$, both coercive fields are exceeded and
the magnetic configuration is converted back to 
QAH.  This QAH state will have the opposite sign of surface magnetization and Hall conductivity compared to the 
starting state (see Fig.~\ref{fig:measurement}(d)). 
The axion state induced on the upsweep part of the hysteresis loop has the same sign of 
incremental voltage as on the downsweep because both the magnetoelectric coefficient 
and the magnetic field have changed sign.  

The absolute value of the TME effect is likely most reliably extracted from the size of the voltage jumps 
that occur near the reversal points in the magnetic hysteresis loop, or equivalently from the currents that would flow through the external circuit to change the capacitor charge if the voltage was held fixed.
The relative correction to quantization $\delta$ is related to the measured voltage jump $\Delta V$ 
\begin{equation}\label{eq:voltage}
 1+\delta  =  \frac{2 C_{\text{TI}} \Phi_0}{eAB^*} \; |\Delta V|,
\end{equation}
where $B^*$ is the magnitude of the field (equal to the top or bottom layer coercive field) at which the jump occurs.  
Expressed in natural units, the voltage jump $|\Delta V| \approx 2.18 (1+\delta) d_{\text{TI}} B^{\ast}/\epsilon_{\text{TI}}$ in units of mV when $[d_{\text{TI}}]=nm$ and $[B^{\ast}]=T$.  
For $\epsilon_{\text{TI}} = 10$, $B^{\ast} = 1T$, $d_{\text{TI}} = 10$ nm, $|\Delta V|$ is therefore around $2.18(1+\delta)$ mV.
Extracting $\delta$ requires knowledge of the capacitance of the TI, 
but this quantity can be measured for the same sample under the same experimental 
conditions by using the total capacitance at $B=0$ in the hysteresis loop and accounting for 
the inverse capacitance contributions from the $h$-BN dielectric layers.  
It is desirable to keep the 
dielectric layers thin in order that this correction, which needs to be estimated from known dielectric constants and sample thicknesses, is relatively small.  Because of this correction, the parts per millions 
accuracy routinely achieved in quantum Hall effect measurements seems out of reach.  
However, this measurement accuracy may be irrelevant since we do not 
expect that the TME is accurately quantized in realistic devices with disorder for fundamental reasons, 
as we now explain. 

\textit{Effect of disorder on the precision of TME quantization---}
In order to begin a discussion of the influence of disorder on TME quantization accuracy, we comment on the physical processes taking place
when the hysteresis loop is transited.  The voltage jumps that take place at negative $B$ on the downsweeps and at 
positive $B$ on the upsweeps occur as the local magnetization direction on either the top or bottom surface is in the process of 
reversing.  These changes in surface magnetization drive anomalous currents that flow through the bulk of the film \cite{Essin2010,mahon2023reconciling}.  On the other hand, the steady rate of voltage change with field
that occurs after the magnetization has completely reversed is due \cite{Essin2010,Pournaghavi2021} to charges that flow between surfaces on the sidewalls.
In both cases, the change in charge distribution across the film is a purely equilibrium phenomenon.  

\begin{figure}[t!]
    \centering
    \includegraphics[width=0.45\textwidth]{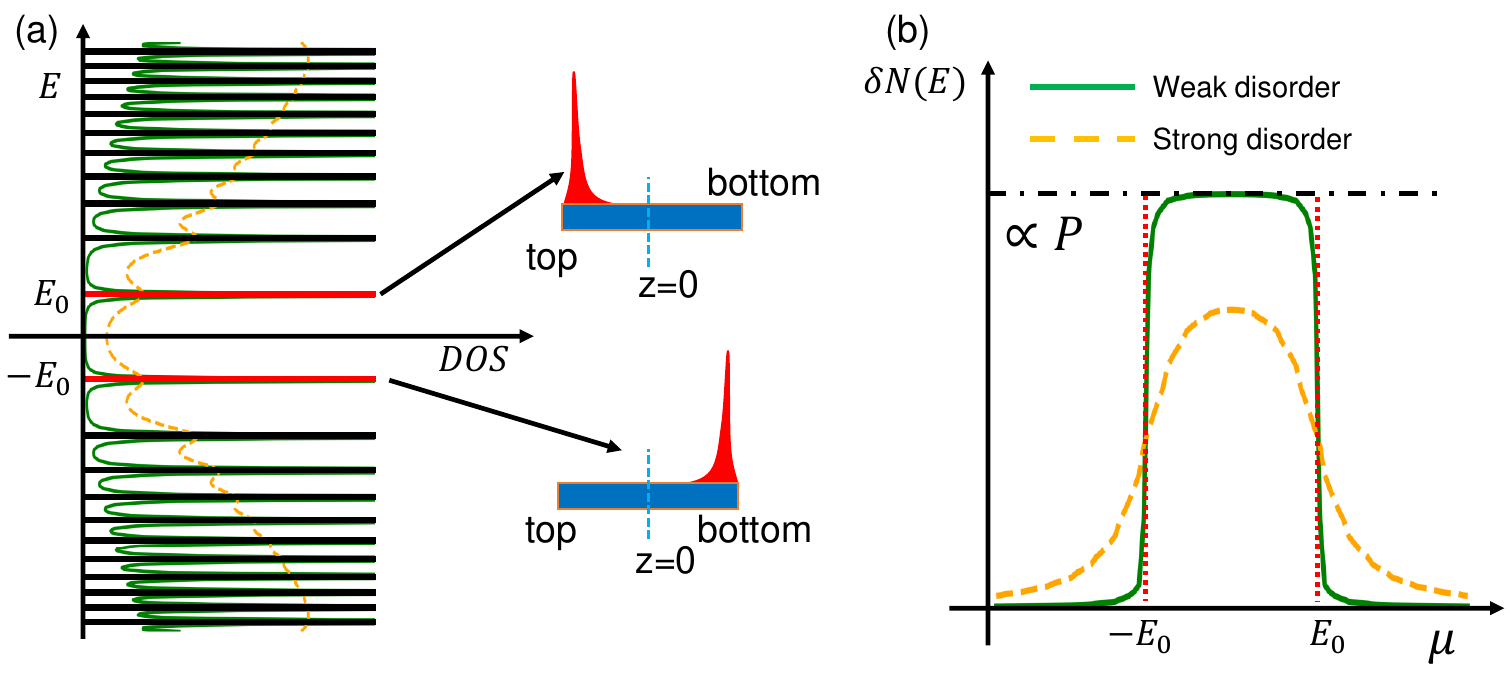}
    \caption{Anomalous Landau levels (LLs) and effect of disorder in axion configured thick films. 
    (a) Effect of disorder on the electronic density of states (DOS): the left panel 
    plots the DOS {\it vs.}~band energy with and without disorder, with the green (orange dashed) curves for weak 
    (strong) disorder and the black (red) lines for disorder-free (anomalous $\mathcal{N}=0$) LLs. 
    Right panel: out-of-surface axion surface magnetic dopant configuration is assumed. 
    (b) Difference between the electronic densities in the top and bottom ($\delta N(E)$) halves of a finite thickness 
    Bi$_{2}$Te$_{3}$ film for weak and strong disorder {\it vs.}~chemical potential.  
    When there is weak disorder there is a charge gap at neutrality and
    $\delta N(E)$ is maximized for chemical potentials in this gap. 
   The size of the energy gap is bounded by that of the
   $\mathcal{N}=0$ LLs in the system with no disorder (denoted $2E_{0}$).
   When there is strong disorder there is no charge gap and there is a single value of $\mu$ at which the film is electrically neutral and $\delta N(E)$ is maximized.
   }
    \label{fig:disorder}
\end{figure}

The influence of disorder on the magnetoelectric response coefficient can be assessed qualitatively by adding 
disorder to a low-energy effective model for nonmagnetic $\mathbb{Z}_{2}$ TI thin films with surface magnetic dopants.  We focus on Bi$_{2}$Te$_{3}$ in the X$_{2}$V$_{3}$ family of layered compounds for which such a model can be constructed using coupled Dirac cones, one associated with the top (bottom) surface of each constituent layer \cite{burkov2011weyl,Lei2020}.  
The parameters of the Dirac cone model can be chosen to provide a semiquantitative 
low-energy description \cite{Lei2020,SupplementalMaterial} of the electronic states in common magnetic and nonmagnetic TIs, and have the advantage 
of simplicity which allows the influence of an external magnetic field to be accounted for by explicit calculation.
In these models, the electric polarization linearly induced by a uniform dc magnetic field aligned with the stacking axis has contributions from the spin-polarized $\mathcal{N}=0$ anomalous Landau level 
subspace only, within which the Hamiltonian is a generalized Su–Schrieffer–Heeger (SSH) \cite{Lei_TME_2023,mahon2023reconciling} 
model with two sites per layer.  In thin films that are nonmagnetic in the bulk, the SSH model
site energies are shifted in opposite directions on opposite surfaces
when the surface magnetic dopants are in an axion (opposing direction) configuration \cite{mahon2023reconciling}, leading to an electric dipole moment.  The axion order produces SSH model localized 
end states with energies that are shifted by $\pm E_{0}$.  (See Fig.~\ref{fig:disorder}(a) in which
the density of states (DOS) versus band energy in the absence of disorder
is plotted using black curves for the $\mathcal{N}\ne 0$ contribution and 
red curves for the $\mathcal{N}=0$ contribution.)

In real materials disorder is always present. 
In TI thin films either spin or charge disorder due to the magnetic dopants or other 
crystal imperfections can induce gapless surface states.  In our Dirac-cone model the
relevant $\mathcal{N}=0$ Landau levels are fully spin polarized and 
these contributions combine indistinguishably.
The DOS in the presence of disorder can be approximated by convolution of 
the disorder-free Landau level spectrum with a finite-width disorder filter.
The DOS calculated using Lorentzian broadening functions \cite{SupplementalMaterial} is illustrated in Fig.~\ref{fig:disorder}(a) using dark green curves for weak disorder and 
orange dashed curves for strong disorder.  
In the presence of strong disorder, the $\mathcal{N}=0$ Landau levels just below the Fermi level, which are localized near the bottom of the film in this illustration, are 
not fully occupied, and the $\mathcal{N}=0$ Landau levels just above the Fermi level, which are localized near the top of the film,
are not fully empty. The difference between the total densities in the top and bottom halves of the thin film
is therefore reduced, as illustrated in Fig.~\ref{fig:disorder}(b).  
The end result is that disorder reduces the magnetoelectric response. 
As shown in Fig.~\ref{fig:disorder}(b), in the case of strong disorder there is no charge gap at neutrality. TIs that have this property cannot have a quantized magnetoelectric response.
Because the TME is an equilibrium and not a 
transport property, precise quantization is not saved by disorder-induced localization even if it is present.
In order for the TME to be accurately quantized, the disorder potentials must be smaller
than $2E_{0}$, the energy difference between $\mathcal{N}=0$ Landau levels.
We remark that since 
the gapless surface states in the disorder-free case are implied by time-reversal symmetry, which
forbids magnetoelectric response, it is magnetic doping that allows the TME to occur.
However, the TME is not quantized unless the doping produces a true spectral gap.

Although surface-state disorder does not save quantization of the TME, it may under some circumstances lead to
interesting and informative noise and frequency dependencies in capacitive voltages measured in stacks similar to those 
in Fig.~\ref{fig:scheme}.  We expect that magnetic noise and hysteresis associated with the single-surface magnetization reversals 
that separate QAH and axion insulator states will be amplified in the capacitive voltages because of the sensitivity of the 
TME to the relative orientations of top and bottom layer magnetizations.  The change in charge distribution across the 
TI is in this case associated with rapid changes in the energy of the surface localized Landau level upon magnetization reversal.
We expect that the time scale for equilibration when the magnetic field is changed within the narrow reversal 
intervals will normally be set by magnetic processes, whereas it will be set 
by electronic processes when the magnetic field is varied across the fixed magnetization axion insulator intervals.
The response to magnetic field in these intervals accommodates changes in the surface state Landau level degeneracy, 
and moves charge between surfaces along the side walls. It is possible that this response will also be slow enough that its 
lag is observable if the surface states are strongly localized, leading to frequency dependence 
in the capacitance.

\textit{Discussion---} 
In this Letter we have discussed measurement of the TME by capacitive detection of polarization response to magnetic field.
It is also interesting to consider the possibility of probing the same quantity by 
measuring the orbital magnetization response to electric fields.  Orbital magnetization can be 
measured directly by detection of the stray magnetic fields they produce \cite{Tschirhart_2021,Grover_2022,Zhou_2023}.
It can also be measured indirectly by measuring Kerr magneto-optical response coefficients, and assuming that the two quantities 
are equivalent up to a separately measurable normalization factor.  Assumptions of this nature are frequently 
made, but can be fully justified only when both quantities are linear in whatever parameter best characterizes time-reversal
symmetry breaking -- in the present case most likely the mean spin polarization of the surface magnetic dopants.
Since the TME is extremely nonlinear in the absence of disorder, appearing instantly for infinitesimal time-reversal 
symmetry breaking, it seems unlikely that Kerr probes can be used for quantitative tests of the accuracy of TME quantization. 

The TME can also be probed indirectly using transport.  In the absence of disorder 
the TME of thin films with small surface magnetizations is equivalent to 
separate half-quantized surface state Hall conductivities of opposite signs on top and bottom film surfaces.
Assuming that the two surfaces conduct in parallel, only one surface will contribute to the Hall conductivity 
when only one surface is magnetized.  Its Hall conductivity can then be extracted by inverting the measured total 
resistivity.  Indeed, experimental observation of a near half-quantized 
 Hall effect has been reported in a thin film with magnetization on one surface only \cite{Mogi_2022,Ralph}. 
This measured surface Hall conductivity implies that charge flows into or out of that surface as magnetic field is varied.
Similarly, measurements of the total QAH effect \cite{Goldhaber-Gordon_QAG_ppm,Dziom_2017} in thin films when the magnetization on the 
two surfaces are parallel can be viewed as providing information that is equivalent to that obtained by measurements in a 
device with only one magnetized surface.  The Hall conductivity is simply twice as large as in the single-magnetic-surface case.
The accuracy of quantization and the finite plateau width of the anomalous Hall effect,
which is critical to identify quantized values, is assisted by disorder which makes states at the 
Fermi level in the two-dimensional bulk irrelevant as long as they are localized.  These surface localized states are 
not irrelevant for the TME in real disordered devices, however, since they are related to the equilibrium 
charge transfer between layers after allowing the system to reequilibrate following any change in the magnetic
field value.  The time scale required for this reequilibration can be estimated by measuring the frequency dependence of the 
capacitance in oscillating magnetic fields.  The TME quantization should be more accurate when the 
magnetic field frequency exceeds the surface equilibration rate.
Ultimately, material platforms capable of hosting the TME that are presently available always have some degree of disorder.
Measurements of the type we propose in this Letter may help provide some 
guidance to materials efforts that are designed to reduce the disorder in 
topological insulator surface states.

\textit{Acknowledgments---}
We acknowledge fruitful discussions with Cui-Zu Chang, Lingjie Zhou, Deyi Zhuo, Xiaoda Liu and Shengwei Chi.
This work was supported by the Robert A. Welch Foundation under Grant Welch F-2112 and by the Simons Foundation. C.M.C. acknowledges support from the Swedish Research Council (VR 2021-046229). 

\bibliography{TME}

\end{document}


\title{Supplementary Material for\\ ``Capacitive Scheme to Detect the Topological Magnetoelectric Effect''}

\date{\today}
\author{Chao Lei$^{1}$}
\email{leichao.ph@gmail.com}
\author{Perry T.~Mahon$^{1}$}
\author{C.~M.~Canali$^{2}$}
\author{A.~H.~MacDonald$^{1}$}
\affiliation{$^{1}$Department of Physics, University of Texas at Austin, Austin, TX 78712, USA}
\affiliation{$^{2}$Department of Physics and Electrical Engineering, Linnaeus University, 392 31  Kalmar, Sweden}

\maketitle

The Hamiltonian used in this work is constructed using Dirac cones that are coupled by hybridizing states:
\begin{equation}\label{Hamiltonian_dirac}
   \hat{H} =  \sum_{\bk_{\perp},ij,\mu\gamma} \Big[\Big( \, 
   (-1)^i  \hbar v_{D}  (\hat{z} \times \bs) \cdot \bk_{\perp} + m_{i} \sigma_z \Big)_{\mu,\gamma} \delta_{ij}  
  + \Delta_{ij}(1-\delta_{ij} )\delta_{\mu,\gamma} \Big] \hat{c}_{\bk_{\perp},i,\mu}^{\dagger} \hat{c}_{\bk_{\perp},j,\gamma}.
\end{equation}
Here we have chosen $\hat{\bm{z}}$ as the stacking axis of constituent quintuple layers, 
$\bk_{\perp}=(k_{x},k_{y})$, $i,j\in\{0,\ldots,2N-1\}$ where $i=2(l-1)$ $(i=2l-1)$ labels states that are associated with the bottom (top) surface of layer $l\in\{1,\ldots,N\}$,
$\mu,\gamma\in\{\uparrow,\downarrow\}$ are spin labels, 
$\Delta_{ij}$ are the intra- and inter-layer hybridization parameters,
$v_{_D}$ is the velocity of an isolated Dirac cone, and $m_{i}$ are mass terms 
that account for exchange coupling to static local magnetic moments. The mass terms $m_{i}$ are related to the magnetizations via the equation $m_{i} = \sum_{l} J_{il} M_{l}$, where $M_{l} = 0$ for non-magnetic layers and $\pm 1$ for magnetic layers; 
the Dirac-like states at the top or bottom surface of layer $l$ 
are coupled to the static magnetization of that layer (via $J_\text{S}$)
and to that of the adjacent layer nearest that surface (via $J_\text{D}$),
and hybridized only to adjacent surfaces, 
with hopping $\Delta_\text{S}$ for opposite surfaces of the same layer and 
$\Delta_\text{D}$ for the closest surface of adjacent layers.
When setting the parameters $J_\text{S} = J_\text{D}$, $M_{1}=-1$ for the even layer, and $M_{N}=1$ for the odd layer, this model describes the non-magnetic topological insulators with a surface magnetization $J_S$.

In the presence of an external magnetic field $\bB = -B \hat{\bm{z}}$, the $\bm{k}\cdot\bm{p}$ model is mapped to an envelope function approximated Hamiltonian that is minimally-coupled to the electromagnetic field: 
$\hbar\bm{k} \rightarrow -i\hbar\bm{\nabla}-(eBx/c)\hat{\bm{y}}$ in the Landau gauge. Here $e$ is the elementary charge unit.
Applying this recipe to
Eq.~(\ref{Hamiltonian_dirac}) and seeking energy eigenvectors of the form $\Psi_{E,q_{y}}(x,y)=e^{iq_{y}y}\Phi_{E,q_{y}}(x)/\sqrt{L_{y}}$, where 
$\Phi_{E,q_{y}}(x)$ has $2N\times2$ components that account for layer and spin label degrees of freedom,
\begin{equation}
\label{eq:Dirac}
    H_D=\hbar v_{_D}  (\hat{z} \times \bs) \cdot \bk_{\perp} \to \hbar \omega_c \begin{pmatrix} 0 & a^{\dagger} \\ a & 0 \end{pmatrix},
\end{equation}
where
$a = \frac{1}{\sqrt{2}}(\tilde{x} + \partial_{\tilde{x}})$ and $a^{\dagger} = \frac{1}{\sqrt{2}}(\tilde{x} - \partial_{\tilde{x}})$ are Landau level raising and lowering (differential)
operators,
$ \omega_c \equiv v_D/\sqrt{2}l_B $, 
$ l_B \equiv \sqrt{\hbar c/eB} $ is the magnetic length, and $ \tilde{x} \equiv l_B q_{y} - x/l_B $.
The density of states in the presence of magnetic field is:
\begin{equation}
    D(E) = \frac{eB}{h} \sum_{l=0}^{\infty} {\delta(E - E_l)},
\end{equation}
where $E_l$ is the Landau level energies.
Considering the disorder effect, the density of states for Landau levels are broadened as:
\begin{equation}
    D(E) = \frac{eB}{\pi h} \sum_{l=0}^{\infty} {\frac{\Gamma}{\Gamma^2 + (E-E_l)^2}}.
\end{equation}
Here we considered a Lorentzian broadening where $\Gamma$ is a parameter specifying the width which is determined by the strength of disorder. The projected integrated density of states to the top and bottom surface defined in the main text are calculated by
\begin{equation}
   N_{t(b)}(E_F) = \sum_{i\le N (i\ge N)} N_i(E_F) = \sum_{i\le N (i\ge N)} \int_{-\infty}^{E_F} D(E) |\Phi_i(E_l)|^2 dE,
\end{equation}
in which $N$ is the thickness of thin films, $E_F$ is the Fermi level, and $i$ labels the Dirac cones locations. The $\delta N(E) \equiv N_t(E_F) - N_b(E_F)$ is then used the characterize the electric polarization.